\documentclass[11pt]{article}
\begin{document}
\title{Logarithmic Conformal Field Theories Near a Boundary}
\author{S. Moghimi-Araghi$^1$ \footnote{e-mail: samanimi@rose.ipm.ac.ir} ,
S. Rouhani$^{1,2}$ \footnote{e-mail: rouhani@karun.ipm.ac.ir}\\
\\
$^1$Department of Physics, Sharif University of Technology,\\
Tehran, P.O.Box: 11365-9161, Iran\\
$^2$Institute for Studies in Theoretical Physics and Mathematics,\\
Tehran, P.O.Box: 19395-5531, Iran}
\maketitle

\begin{abstract}
We consider logarithmic conformal field theories near a boundary and derive the
general form of one and two point functions. We obtain results for arbitrary 
and two dimensions. Application to two dimensional magnetohydrodynamics is discussed.
\end{abstract}

\section{Introduction}

Gurarie has pointed out \cite{Gur} that it is possible to construct consistent 
conformal field theories which have logarithmic terms in their correlation functions.
In these theories representation of the Virasoro algebra is not diagonalizable.
Logarithmic conformal field theories (LCFTs) contain logarithmic operators 
which have logarithms as well as powers in their operator product expansion.
Such operators do not appear in unitary CFTs. However this does not mean that they are not
relevant. By now a large number of examples of the applications of such theories 
have been found. For some examples see citations in \cite{Flohr sing}.
Extensive work has been done on the structure of LCFTs \cite{Flohr sing,GK,RAK}, although a lot
remains to be worked out.

Conformal invariance in physical systems arises at a fixed point. It is also known that
critical behavior near boundaries is affected by the geometry of the boundary.
Indeed correlators of conformal field theories near boundaries have extra 
structure which gives rise to the surface critical behavior \cite{Cardy}. Conformal
theories of turbulence near a boundary were discussed by Chung et al
\cite{Koreans}.

In this paper we investigate such theories near a boundary. For definiteness,
we consider a semi-infinite $d$-dimensional system bounded by a ($d-1$)-dimensional
plane surface. The results of this paper fall into two classes. For arbitrary 
dimensions $d$, we show that conformal invariance determines the one-point functions
up to some constants and restricts the form of two-point functions up to some unknown
functions of a single scaling variable. In two-dimensions our result is much stronger.
We show that in this case the n-point functions in the semi-infinite geometry satisfy
the same equations as the 2n-point functions in the bulk. In the end, we apply 
the results derived in sections II and III, to the 2D MHD problem and find
the mean values of velocity and magnetic field.   

\section{Arbitrary Dimensions}

In logarithmic conformal field theories, there exist blocks of fields which
constitute a non-diagonalizable representation of the conformal group. Under
a conformal transformation ${\bf r}\rightarrow{\bf r}^{\prime}=f({\bf {r}})$ 
fields of a block trasmform as \cite{GK}:

\begin{equation}
\Phi^{\prime}({\bf r}^{\prime})
=\left|\frac {\partial{{\bf r}^{\prime}}}
{\partial{\bf r}}\right|^{T}\Phi({\bf r})
\end{equation}
where 
\begin{equation}
  \Phi=\def\temp{\multicolumn{1}{c|}{1}}
  \pmatrix{\phi_{1}\cr \phi_{2} \cr
  \vdots\cr
  \phi_{n}\cr} 
\end{equation}
and T is Jordanian matrix
\begin{equation}
 T = \pmatrix{-\frac {\Delta_{\phi}}{d}&0&\ldots&0\cr 1&-\frac {\Delta_{\phi}}{d}&\ldots&\vdots\cr
 0&1& \ldots&0 \cr
 \vdots&\ldots&\ddots&0 \cr
 0&\ldots&1&-\frac {\Delta_{\phi}}{d}\cr}~. 
\end{equation}

Here $d$ is dimension of space-time and $\Delta_{\phi}$ is conformal
dimension of $\phi_{i}$'s. As an example, for a two-dimensional Jordan
cell  $\Phi=\phi_{1}$ and $\Psi=\phi_{2}$, equation (1) reduces to:
\begin{eqnarray}
\Phi^{\prime}({\bf{r^{\prime}}})&=&
\left|\frac {\partial{{\bf{r^{\prime}}}}}
{\partial{\bf{r}}}\right|^{-\frac{\Delta_{\phi}}{d}}\Phi({\bf{r}})\nonumber\\
\Psi^{\prime}({\bf{r^{\prime}}})&=&
\left|\frac {\partial{{\bf{r^{\prime}}}}}{\partial{\bf{r}}}
\right|^{-\frac{\Delta_{\phi}}{d}}\left(\Psi({\bf{r}})
+\log|\frac {\partial{\bf{r^{\prime}}}}{\partial{\bf{r}}}
|\Phi(\bf{r})\right).
\end{eqnarray}

In $d$ dimensions, invariance under conformal transformations determines
two- and three-point functions up to some constants and four-point 
functions are determined up to functions of crossing ratios. These invariances exist only
if the geometry of the problem is not changed by the transformations. So if there
is a boundary in the problem, correlation functions are invariant only under 
transformations which preserve the boundary. In this case we have less
symmetries and correlation functions are not easily determined~.
To be more specific, we investigate correlation functions of a two-dimensional
Jordanian cell with a simple boundary~--~a $(d-1)$-dimensional hyper-plane.
The symmetry group that preserves the geometry is made up of translations
along the boundary, rotations in the boundary, dilatation and special
conformal transformation along the boundary. We take as the boundary,
the hyper-plane $y=0$.

Consider the one-point functions $\left<\Phi({\bf{r}})\right>=f_{1}(\bf{r})$~,
$\left<\Psi(\bf{r})\right>=f_{2}(\bf{r})$ near the boundary. Invariance under translations
along the surface implies that $f_{1}$ and $f_{2}$ depend only on y. Also under an
infinitesimal dilatation transformation:
\begin{equation}
\bf{r^{\prime}}=\bf{r}+\epsilon\bf{r}
\end{equation}
we must have:
\begin{eqnarray}
f_{1}(y)&=&(1+\epsilon)^{\Delta_{\phi}}f_{1}(y^{\prime})~,\nonumber\\
f_{2}(y)&=&(1+\epsilon)^{\Delta_{\phi}}\left(f_{2}(y^{\prime})
+\log(1+\epsilon)^{d}f(y^{\prime})\right)~.
\end{eqnarray}

Expanding these equations to the first order in $\epsilon$, $f_{1}$ and $f_{2}$
satisfy the differential equations: 
\begin{eqnarray}
y\frac {\partial{f_{1}}}{\partial{y}}+\Delta_{\phi}f_{1}&=&0\nonumber\\
y\frac {\partial{f_{2}}}{\partial{y}}+\Delta_{\phi}f_{2}+d\:f&=&0
\end{eqnarray}
which yield:
\begin{eqnarray}
\left<\Phi(\bf{r})\right>&=&\frac {C_{1}}{y^{\Delta_\phi}} \nonumber\\
\left<\Psi(\bf{r})\right>&=&\frac {1}{y^{\Delta_\phi}}(C_{2}-d\:C_{1}\log{y}) 
\end{eqnarray}

Invariance under rotation and special conformal transformations reveals no
other constraint on these one-point functions.

Now consider the two-point function $G_{1}({\bf{r}}_{1},{\bf r}_{2})
=\left<\Phi({\bf r}_{1})\Phi({\bf r}_{2})\right>$~.
The points ${\bf r}_{1}$ and ${\bf r}_{2}$ lie in a unique plane perpendicular to the
surface. Within this plane we can specify the points by coordinates $(x_{1},y_{1})$, $(x_{2},y_{2})$.
By translational invariance parallel to the surface, $G_{1}$ depends on $x_{1}-x_{2}$,
$y_{1}$ and $y_{2}$. Invariance under dilatation implies:
\begin{equation}
G_{1}(x_{1}-x_{2},y_{1},y_{2})=(1+\epsilon)^{\Delta_{\phi}}
(1+\epsilon)^{\Delta_{\phi}}G(x_{1}^{\prime}-x_{2}^{\prime}
,y_{1}^{\prime},y_{2}^{\prime})~.
\end{equation}
An infinitesimal special conformal transformation along x is 
\begin{eqnarray}
x^{\prime}&=&x+\epsilon(x^{2}-y^{2})\nonumber\\
y^{\prime}&=&y+2\epsilon xy,
\end{eqnarray}
and under such a transformation we have 
\begin{equation}
G_{1}(x_{1}-x_{2},y_{1},y_{2})=
(1+2\epsilon x_{1})^{\Delta_{\phi}}(1+2\epsilon x_{2})^{\Delta_{\phi}}G(x_{1}^{\prime}-x_{2}^{\prime},y_{1}^{\prime},y_{2}^{\prime})~.
\end{equation}
By expanding equations (9) and (11) to the first order in $\epsilon$, we arrive at~:
\begin{eqnarray}
u \frac {\partial{G_{1}}}{\partial{u}}
+ y_{1} \frac{\partial{G_{1}}}{\partial{y_{1}}}
+ y_{2} \frac{\partial{G_{1}}}{\partial{y_{2}}}
+ 2{\Delta}_{\phi}G_{1}&=&0\nonumber\\
({y_{1}^{2}}-{y_{2}^{2}})
\frac {\partial{G_{1}}}{\partial{u}} +
u\left(y_{1} \frac{\partial{G_{1}}}{\partial{y_{1}}}
- {y_{2}} \frac {\partial{G_{1}}}{\partial{y_{2}}}\right)&=&0
\end{eqnarray}
in which $u=x_{1}-x_{2}$. The first equation states that $G_{1}$ is a homogeneous
function of dimension $2\Delta_{\phi}$:
\begin{equation}
G_{1}=\frac {1}{(u)^{2\Delta_{\phi}}}g_{1}(\alpha,\beta)
\end{equation}
where $\alpha={y_{1}}/{u}$ and $\beta={y_{2}}/{u}$. Substituting
this in the second line of equation (12) one finds \cite{Cardy}
\begin{equation}
\left[\alpha + \frac {\alpha}{\alpha^{2}-\beta^{2}}\right]\frac {\partial{g_{1}}}{\partial{\alpha}}+\left[\beta + \frac {\beta}{\beta^{2}-\alpha^{2}}\right]
\frac {\partial{g_{1}}}{\partial{\beta}}+2\Delta_{\phi}g_{1}=0~.
\end{equation}
The general solution of this equation is:
\begin{equation}
g_{1}(\alpha,\beta)=\frac {1}{(\alpha\beta)^{\Delta_{\phi}}}h_{1}\left(\frac {1+(\alpha-\beta)^{2}}{\alpha\beta}\right)~.
\end{equation}
So the two-point correlation function is:
\begin{equation}
\left<{\Phi}({\bf r}_{1})\Phi({\bf r}_{2})\right>=\frac {1}{(y_{1}y_{2})^{\Delta_{\phi}}}h_{1}\left(\frac {(x_{1}-x_{2})^{2}+(y_{1}-y_{2})^{2}}{y_{1}y_{2}}\right)~.
\end{equation}
We have two other two-point functions, $G_{2}({\bf r}_{1},{\bf r}_{2})
=\left<\Phi({\bf r}_{1})\Psi({\bf r}_{2})\right>$ and
$G_{3}({\bf r}_{1},{\bf r}_{2})=\left<\Psi({\bf r}_{1})
\Psi({\bf r}_{2})\right>$.
We can follow similar steps for these two and the result is 
\begin{eqnarray}
G_{2}({\bf r}_{1},{\bf r}_{2})&=&
\frac {1}{(y_{1}y_{2})^{\Delta_{\phi}}}
\left[h_{2}(\eta)-d\:\log{y_{2}}\:\:h_{1}(\eta)\right]\nonumber\\
G_{3}({\bf r}_{1},{\bf r}_{2})&=&
\frac {1}{(y_{1}y_{2})^{\Delta_{\phi}}}\left[h_{3}(\eta)-d\:\log{y_{1}y_{2}}
\:\:h_{2}(\eta)
+d^{2}\:\log{y_{1}}\log{y_{2}}\:\:h_{1}(\eta)\right]
\end{eqnarray}
where $\eta=[(x_{1}-x_{2})^{2}+(y_{1}-y_{2})^{2}]/{y_{1}y_{2}}$.

Far from the boundary, the effect of boundary becomes negligible and we must recover
the bulk two-point functions: 
\begin{eqnarray}
\left<\Phi({\bf r}_{1})\Phi({\bf r}_{2})\right>&=&0\nonumber\\
\left<\Phi({\bf r}_{1})\Psi({\bf r}_{2})\right>&=&
\frac {a}{r^{2\Delta_{\phi}}}\nonumber\\
\left<\Psi({\bf r}_{1})\Psi({\bf r}_{2})\right>&=&
\frac {1}{r^{2\Delta_{\phi}}}(b-d\:a\log{r})
\end{eqnarray}
where $r=|{\bf r}_{1}-{\bf r}_{2}|$ and $a, b$ are arbitrary constants. These
equations were first derived by \cite {Ts} in two dimensions and were
generalized to d-dimensions by \cite {GK}.

To go far from the boundary one must let $y_{1}$ and $y_{2}$ tend to infinity
keeping $y_{1}-y_{2}$ and $x_{1}-x_{2}$ finite. This means letting $\eta$ tend to zero .
So we can find $h_{1}$ , $h_{2}$ , $h_{3}$ in the limit $\eta\rightarrow{0}$~:
\begin{eqnarray}
h_{1}(\eta)&=&\frac {1}{\eta^{\Delta_{\phi}}}\left(\frac {4\frac {a}{d}}{\log{\eta}}+\frac {C_{1}}{(\log{\eta})^{2}}+\ldots\right)\nonumber\\
h_{2}(\eta)&=&\frac {1}{\eta^{\Delta_{\phi}}}\left(-a+\frac {C_{2}}{(\log{\eta})}+\ldots\right)\nonumber\\
h_{3}(\eta)&=&\frac {1}{\eta^{\Delta_{\phi}}}\left(b-d\:C_{2}-\frac {d^{2}}{4}C_{1}+\ldots\right)
\end{eqnarray}
Here $C_{1}$ and $C_{2}$ are arbitrary constants. On the other hand the behaviour
of these functions when $\eta$ tends to infinity, determines the surface behaviour
of correlation functions (To investigate the surface behaviour one must let $x_{1}-x_{2}$ tend to infinity while keeping $y_{1}$ and $y_{2}$
finite).

To find the surface exponents one should know the behaviour of these functions in this limit~. 
This requires a knowledge of the differential equations governing $h(\eta)$
which requires details of the structure of conformal field theory. 

\section{Two Dimensions}

In two dimensions, however, conformal group is an infinite dimensional group
and any analytic function from the plane to itself is a conformal transformation~.
In LCFT's, under a conformal transformation $z\:\rightarrow\:w(z)$ , $ \bar{z}\:\rightarrow\:w(\bar{z})$
primary fields of a Jordanian cell, $\Phi_{i}$'s, transform as \cite{RAK}:
\begin{equation}
\Phi_{i}(z,\bar{z})\rightarrow \Phi_{i}(z,\bar{z})+\left[\alpha^{\prime}(z)
\Delta_{i}^{j}+\delta_{i}^{j}\alpha(z)\frac {\partial}{\partial{z}}+
\overline{\alpha^{\prime}(z)}\:
{\bar{\Delta}}_{i}^{j}+\delta_{i}^{j}\overline{\alpha(z)}
\frac{\partial}{\partial{\bar{z}}}\right]
\Phi_{j}(z,\bar{z})
\end{equation}
The effect of a small transformation may be expressed in terms of correlation functions
of the fields with the energy-momentum tensor.
In complex coordinates, there are two non-zero components, namely $T(z)=T_{zz}(z)$
and $\bar{T}(\bar{z})=T_{\bar{z}\:\bar{z}}(\bar{z})$.
In terms of these two we have:
\begin{eqnarray}
\frac 1{2\pi i}\oint_{c}dz\:\alpha(z)\left<T(z)\Phi_{i_{1}}(z_{1},\bar{z_{1}})\ldots\right>
-\frac 1{2\pi i} \oint_{c}d\bar{z}\:\overline{\alpha(z)}\left<\bar{T}
(\bar{z})\Phi_{i_{1}}(z_{1},{\bar{z}}_{1})\ldots\right>\nonumber\\
=\sum_{k}\sum_{j_{k}}\left[\alpha^{\prime}(z_{k})
\Delta^{j_{k}}_{i_{k}}+\delta^{j_{k}}_{i_{k}}\alpha(z_{k})\frac {\partial}{\partial{z_{k}}}+
\overline{\alpha^{\prime}(z_{k})}{\bar{\Delta}}^{j_{k}}_{i_{k}}
+\delta^{j_{k}}_{i_{k}}\overline{\alpha(z_{k})}
\frac {\partial}{\partial{\bar{z}}_{k}}\right]
\left<\Phi_{j_{1}}(z_{1},{\bar{z}}_{1})\ldots\right>
\end{eqnarray}
where $c$ is an arbitrary contour containing all the points $z_{k}$ and $k$ is
summed over the fields in the correlators and $j_{k}$ is summed over the
Jordanian cell containing $\Phi_{i_{k}}$. In the
absence of a boundary, $\alpha(z)$ is an arbitrary analytic function, thus 
$\alpha$ and $\bar{\alpha}$ can be assumed independent. As a consequence
the $z$ and $\bar{z}$ dependence in equation (21) separates:
\begin{equation}
\frac 1{2\pi i}\oint_{c}dz\:\alpha(z)\left<T(z)\Phi_{i_{1}}(z_{1},{\bar{z}}_{1})
\ldots\right>
=\sum_{k}\sum_{j_{k}}\left[\alpha^{\prime}(z_{k})\Delta^{j_{k}}_{i_{k}}+
\delta^{j_{k}}_{i_{k}}\alpha(z_{k}) \frac {\partial}{\partial{z_{k}}}\right]
\left<\Phi_{j_{1}}(z_{1},{\bar{z}}_{1})\ldots\right>
\end{equation}
with a similar equation for $\left<\bar{T}\Phi\ldots\right>$. Using Cauchy's theorem
the correlation function $\left<T\Phi\ldots\right>$ can be expressed in terms of linear
differential operators on $\left<\Phi\ldots\right>$ :
\begin{equation}
\left<T(z)\Phi_{i_{1}}(z_{1},{\bar{z}}_{1})\ldots\right>=\sum_{k}\sum_{j_{k}}
\left[\frac {\Delta^{j_{k}}_{i_{k}}}{(z-z_{k})^{2}}
+\frac {\delta^{j_{k}}_{i_{k}}}{(z-z_{k})}\frac {\partial}{\partial{z_{k}}}\right]
\left<\Phi_{j_{1}}(z_{1},{\bar{z}}_{1})\ldots\right>\:.
\end{equation}
In the presence of a boundary, transformations are restricted to those which leave
the boundary invariant, therefore the function $\alpha$ and $\bar{\alpha}$
are no longer independent. Let us take as boundary the real axis, thus the operators
$\Phi_{i_{k}}$ are defined in the upper half-plane $Im(z)>0$.
The transformations which leave this boundary invariant, are the real analytic functions,
i.e $\overline{\alpha(z)}=\alpha(\bar{z})$.
As result of this the separation of $z$ and $\bar{z}$ can not take place.

However, as Cardy has shown \cite{Cardy}, we may extend the definition of $T(z)$ into the lower
half-plane by:
\begin{equation}
T(z):=\bar{T}(z)\:\:\:\:\:\: Im(z)<0
\end{equation}
and relable ${\bar{z}}_{k}=z^{\prime}_{k}$. Changing variable $\bar{z}\rightarrow z$
in the second integral of equation (21) and using $\overline{\alpha(z)}=\alpha(\bar{z})$ ,
it becomes:
\begin{eqnarray}
\frac 1{2\pi i}{\oint}_{c}dz\:\alpha(z)\left<T(z)\Phi_{i_{1}}(z_{1},
z^{\prime}_{1})\ldots\right>+\frac 1{2\pi i}\oint_{\:\bar{c}}
dz\:\alpha(z)\left<T(z){\Phi}_{i_{1}}(z_{1},z^{\prime}_{1})\ldots\right>
\nonumber\\
=\sum_{k}\sum_{j_{k}}\left[{\alpha}^{\prime}(z_{k})\Delta^{j_{k}}_{i_{k}}+
\delta^{j_{k}}_{i_{k}}\alpha(z_{k})\frac {\partial}{\partial{z_{k}}}+
{\alpha}^{\prime}(z^{\prime}_{k}){\bar{\Delta}}^{j_{k}}_{i_{k}}
+\alpha(z^{\prime}_{k})\frac {\partial}{\partial{z^{\prime}_{k}}}\right]
\left<\Phi_{j_{1}}(z_{1},z^{\prime}_{1})\ldots\right> ,
\end{eqnarray}
where $\bar{c}$ is a contour in the lower half-plane\cite{Cardy}.

The left-hand side of equation (25) can be written as one integral around a large
contour containing all the points $z_{k}$ and $z^{\prime}_{k}$ if we have $T=\bar{T}$ on the boundary.
This condition is equivalent to the 
condition $T_{xy}=0$ in Cartesian coordinates which means that there is no flux 
of energy across the boundary. Now that the left-hand side of equation (25) is an integral,
one can use Cauchy's theorem to get:
\begin{eqnarray}
\left<T(z)\Phi_{i_{1}}(z_{1},{z'}_{1})\ldots\right>=
\:\:\:\:\:\:\:\:\:\:\:\:\:\:\:\:\:\:\:\:\:\:\:\:\:\:\:\:\:\:\:\:\:\:\:\:\:\:\:\:\:
\:\:\:\:\:\:\:\:\:\:\:\:\:\:\:\:\:\:\:\:\:\:\:\:\:\:\:\:\:\:\:\:\:\:\:\:\:\:\:\nonumber\\
\sum_{k}\sum_{j_{k}}\left[\frac {\Delta^{i_{k}}_{j_{k}}}{(z-z_{k})^{2}}+
\frac {\delta^{i_{k}}_{j_{k}}}{(z-z_{k})}\frac {\partial}{\partial{z_{k}}}
+\frac {{\bar{\Delta}}^{j_{k}}_{i_{k}}}{(z-z^{\prime}_{k})^{2}}
+\frac {\delta^{j_{k}}_{i_{k}}}{(z-z^{\prime}_{k})}
\frac {\partial}{\partial{z^{\prime}_{k}}}\right]
\left<\Phi_{j_{1}}(z_{1},z^{\prime}_{1})\ldots\right>\:.
\end{eqnarray}
Comparing equations (26) and (23), we observe that in presence of a boundary, the
correlation function
$\left<\Phi_{i_{1}}(z_{1},\bar{z}_{1})
\ldots\Phi_{i_{2n}}(z_{2n},\bar{z}_{2n})\right>$
regarded as a function of $(z_{1},\ldots,z_{n},\newline
{\bar{z}}_{1},\ldots,{\bar{z}}_{n})$
satisfies the same differential equations as that of the bulk correlation
function
$\left<\Phi_{i_{1}}(z_{1},{\bar{z}}_{1})\ldots\Phi_{i_{2n}}(z_{2n},{\bar{z}}_{2n})\right>$ 
as a function of $z_{1},\ldots,z_{2n}$. Note that the conformal dimension of fields
$\Phi_{i_{n+1}}$ to $\Phi_{i_{2n}}$
is $\bar{\Delta}$.

As an example we calculate the one point function of such fields.
Consider a $2 \times 2$ scalar Jordanian cell, composed of the fields $\Phi$ and $\Psi$
($\Psi$ is the logarithmic partner).
Conformal invariance of
correlation functions means that acting with the set ${L_{0},L_{\pm{1}}}$
on the correlation functions yields zero.
Solving the equations obtained, one arrives at:
\begin{eqnarray}
\left<\Phi(z,\bar{z})\right>&=&\frac{c}{(z-\bar{z})^{2\Delta}}\:,\nonumber\\
\left<\Psi(z,\bar{z})\right>&=&\frac 1{(z-\bar{z})^{2\Delta}}
\left[c^{\prime}-2c\log{(z-\bar{z})}\right].
\end{eqnarray}
which are the same as two-point functions without the boundary. Also note that these results are
consistent with the results obtained for general dimensions in the previous section (equation (8)).

Now consider the two-point correlation functions near the boundary. Again 
the differential equations which are satisfied by these two-point correlation
functions are the same as the equations satisfied by four-point correlation function in the bulk
and so the result is:
\begin{eqnarray}
\left<\Phi(z_{1})\Phi(z_{2})\right>&=&u^{2\Delta}f_{1}(v)\:\:\:\:\:\:\:\:\:\:\:\
\:\:\:\:\:\:\:\:\:\:\:\:\:\:\:\:\:\:\:\:\:\:\:\:\:\:\:\:\:\:\:\:\:\nonumber
\end{eqnarray}
\begin{eqnarray}
\left<\Phi(z_{1})\Psi(z_{2})\right>&=&u^{2\Delta}\left(f_{2}(v)-2\log(z_{2}-\bar{z_{2}})f_{1}(v)\right)\nonumber
\end{eqnarray}
\begin{eqnarray}
\:\:\:\:\:\:\:\:\:\:\:\:\:\:\:\:\:\:\:\:\:\:\:\:\:\:\left<\Psi(z_{1})\Psi(z_{2})\right>~=~
u^{2\Delta}(f_{3}(v)-2\log\left[(z_{1}-{\bar{z}}_{1})
(z_{2}-{\bar{z}}_{2})\right]f_{2}(v)
\nonumber\\ \:\:\:\:\:\:\:\:\:\:\:\:\:\:\:\:\:\:\:\:\:\:\:\:\:\:
+4\log(z_{1}-{\bar{z}}_{1})\log(z_{2}-{\bar{z}}_{2})f_{1}(v))
\end{eqnarray}
where $u={(z_{1}-{\bar{z}}_{1})(z_{2}-{\bar{z}}_{2})}/
{(z_{1}-z_{2})({\bar{z}}_{1}-{\bar{z}}_{2})
(z_{1}-{\bar{z}}_{2})({\bar{z}}_{1}-z_{2})}$ and
\newline
$v=(z_{1}-z_{2})({\bar{z}}_{1}-{\bar{z}}_{2})/(z_{1}-{\bar{z}}_{1})
(z_{2}-{\bar{z}}_{2})\:$ and
$f_{1},f_{2},f_{3}$ are arbitrary functions (Again compare these results with
the equations (16) and (17) derived for general dimensions).

If a singular vector is found in such a theory, one can find some differential
equations which are satisfied by $f_{1},f_{2},f_{3}$ (usually they are hypergeometric
ones) and hence the correlation functions will be determined completely. Finding
the correlation functions one can investigate the surface behaviour of the 
theory. In a successing paper we will consider such a theory.     

\section{Application to MHD}

The incompressible two-dimensional magnetohydrodynamic system has two
independent dynamical variables, the velocity stream function $\phi$ and
the magnetic-flux function $\psi$. They obey the equations
\begin{eqnarray}
\frac {\partial{w}}{\partial{t}}&=&-\epsilon_{\alpha\beta}\partial_{\alpha}{\phi}\partial_{\beta}{w}+\epsilon_{\alpha\beta}\partial_{\alpha}\psi\partial_{\beta}{J}+\mu\nabla^{2}w~\nonumber\\
\frac {\partial{\psi}}{\partial{t}}&=&-\epsilon_{\alpha\beta}\partial_{\alpha}{\phi}\partial_{\beta}{\psi}+\eta J~,
\end{eqnarray}
where the vorticity $w=\nabla^{2}\phi$ and the current $J=\nabla^{2}\psi$ and $\mu$ and $\eta$ are viscosity and molecular resistivity. The velocity and
magnetic field are given in terms of $\phi$ and $\psi$:
\begin{eqnarray}
V_{\alpha}&=&\epsilon_{\alpha\beta}\partial_{\beta}{\phi}~\nonumber\\
B_{\alpha}&=&\epsilon_{\alpha\beta}\partial_{\beta}{\psi}~,
\end{eqnarray}
where $\epsilon_{\alpha\beta}$ is the totally antisymmetric tensor with $\epsilon_{12}=1$.
It has been argued that the Alf'ven effect implies that $\phi$ and $\psi$ should 
have equal scaling dimension which naturally leads to LCFT's~\cite{RR,Flohr MHD}.

We consider this system near a boundary and calculate the mean values of velocity
and magnetic field. As we have derived in the last section the one-point function of the fields $\phi$ 
and $\psi$ are given by equation (27),
so for velocity we have:
\begin{eqnarray}
\left<V_{x}(x,y)\right>&=&\partial_{y}\left<\phi(x,y)\right>=-\frac {2\Delta C}{y^{2\Delta+1}}~\nonumber\\
\left<V_{y}(x,y)\right>&=&-\partial_{x}\left<\phi(x,y)\right>=0~,
\end{eqnarray}
and for magnetic field:
\begin{eqnarray} 
\left<B_{x}(x,y)\right>&=&\partial_{y}\left<\psi(x,y)\right>=-\frac {2\Delta}{y^{2\Delta}}\left[(C^{\prime}+2C)-2C\log y\right]~\nonumber\\
\left<B_{y}(x,y)\right>&=&-\partial_{x}\left<\psi(x,y)\right>=0~.
\end{eqnarray}

A specific model is proposed by Rahimi-Tabar and Rouhani~\cite{RR} with $\Delta=\frac {-5}{7}$~. 
This theory seems to be unphysical at first sight, because V and B grow
large far from boundary. However, they acquire physical meaning when this model
is regularized, for example by attaching a value to the velocity field at the boundary.

Other boundaries such as strip and circle can be readily investigated by
proper transformations. For example for the strip geometry with size $L$ one obtains: 

\begin{eqnarray}
\left<\phi(x,y)\right>&=&\left(\frac {\pi}{L}
\right)^{2\Delta}\frac {C}{(\sin{\frac {\pi}{L}}{y})^{2\Delta}}\nonumber\\
\left<\psi(x,y)\right>&=&\left(\frac {\pi}{L}\right)^{2\Delta}
\frac {1}{(\sin{\frac {\pi}{L}}{y})^{2\Delta}}
\left(C^{\prime}+2C\log{\frac {\pi}{L}}-2C\log\sin({\frac {\pi}{L}}{y})
\right).
\end{eqnarray}

Further development of LCFT, such as complete calculation of the four point functions, 
is necessary before some interesting questions such as the possible 
set of surface critical indices can be determined. Work in this direction is in progress.

{\large {\bf Acknowledgement} }

We would like to thank M. R. Rahimi-Tabar and M. Sa'addat for helpful comments
and advice on this paper.

\end{document}